\documentclass{iopart}
\usepackage{iopams}

\expandafter\let\csname equation*\endcsname\relax
\expandafter\let\csname endequation*\endcsname\relax
\usepackage{amsmath}
\usepackage{graphicx}

\graphicspath{{figs/}}
\usepackage[colorlinks]{hyperref}

\usepackage{xcolor}
\usepackage{amssymb}
\usepackage{tikz}
\usepackage{adjustbox}

\usetikzlibrary{shapes,arrows,positioning}

\tikzstyle{block} = [draw, rectangle, fill=yellow!20,
    minimum height=2em, minimum width=3em, inner sep=0.5em]
\tikzstyle{sum} = [draw, circle, inner sep=0pt, minimum size=0.6em]
\tikzstyle{junction} = [draw, circle, fill, inner sep=0pt, minimum size=1mm]
\tikzstyle{coord} = [coordinate]
\tikzstyle{connector} = [->,thick]
\tikzstyle{tag} = [node distance=1mm]

\begin{document}

\title[]{Simultaneous feedback control of toroidal magnetic field and plasma current on MST using advanced programmable power supplies}

\author{I. R. Goumiri$^{1, 2}$, K. J. McCollam$^2$,  A. A. Squitieri$^2$, D. J. Holly$^2$, J. S. Sarff$^2$ and S. P. Leblanc$^3$.}
\address{$^1$ Lawrence Livermore National Laboratory, Livermore, CA 94551, USA}
\address{$^2$ Department of Physics, University of Wisconsin-Madison, Madison, WI 53719, USA}
\address{$^3$ Unaffiliated}

\ead{goumiri1@llnl.gov}
\vspace{10pt}
\begin{indented}
\item[]
\end{indented}

\begin{abstract}

Programmable control of the inductive electric field enables advanced operations of reversed-field pinch (RFP) plasmas in the
Madison Symmetric Torus (MST)
device and further develops the technical basis for ohmically heated fusion RFP plasmas.
MST's poloidal and toroidal magnetic fields ($B_\text{p}$ and $B_\text{t}$) can be sourced by
programmable power supplies (PPSs)
based on integrated-gate bipolar transistors (IGBT).
In order to provide real-time simultaneous control of both $B_\text{p}$ and $B_\text{t}$ circuits, a time-independent integrated model is developed. The actuators considered for the control are the $B_\text{p}$ and $B_\text{t}$ primary currents produced by the PPSs. The control system goal will be tracking two particular demand quantities
that can be measured at the plasma surface ($r=a$): 
the plasma current, 
$I_\text{p} \sim B_\text{p}(a)$, and the RFP reversal parameter, $F\sim B_\text{t}(a)/\Phi$, where $\Phi$ is the toroidal flux in the plasma.
The edge safety factor, $q(a)\propto B_t(a)$, tends to track $F$ but not identically.
To understand the responses of $I_\text{p}$ and $F$ to the actuators and to enable systematic design of control algorithms, dedicated experiments are run in which the actuators are modulated, and a linearized dynamic data-driven model is generated using a system identification method. 
We perform a series of initial real-time experiments to test the designed feedback controllers and validate the derived model predictions. The feedback controllers show systematic improvements over simpler feedforward controllers.

\end{abstract}



\section{Introduction}
\label{intro}

The reversed field pinch (RFP) is a toroidal magnetic confinement configuration that has the potential to achieve an ohmically heated and inductively sustained steady-state fusion plasma. In contrast to the tokamak configuration, the RFP is magnetized primarily by plasma current. The magnetic equilibrium has low safety factor, $|q(r)|\lesssim a/2R_0$, allowing the current density and ohmic heating to be much larger than for a tokamak plasma of the same size and magnetic field strength~\cite{Christiansen82}. Furthermore, the RFP plasma exhibits magnetic relaxation that is subject to conservation of magnetic helicity~\cite{Taylor86,Ji95}. This allows the possibility for using AC magnetic helicity injection, also called oscillating field current drive
 (OFCD),
to sustain a steady-state plasma current using purely AC inductive loop voltages~\cite{Bevir85,McCollam06}. An ohmically heated and inductively sustained plasma could greatly simplify a toroidal magnetic fusion reactor by eliminating the need for auxiliary heating and non-inductive current drive.

Key to achieving a steady-state, ohmically heated RFP will be advanced, programmable control of the poloidal and toroidal field magnets and their power supplies. Programmable power supplies are used in many fusion experiments, but the RFP has the special challenge of large power flow between the poloidal and toroidal magnetic field circuits via nonlinear relaxation processes regulated by the plasma. This is particularly acute with OFCD, where megawatts of reactive power oscillate between the circuits and regulated by the plasma~\cite{McCollam10}. Precise phase control of the AC toroidal and poloidal inductive loop voltages is essential for OFCD.

The relaxation process appearing in RFP plasmas happens through nonlinear interactions of tearing instabilities that cause magnetic turbulence, which tends to degrade energy confinement. This turbulence decreases with increasing plasma current and commensurate higher plasma temperature, but it is still uncertain if
energy confinement
scaling will be 
sufficient to reach 
ohmic ignition. Inductive pulsed parallel current drive 
(PPCD)
control shows that the RFP plasma can achieve similar confinement to a tokamak of the same size and magnetic field strength when tearing instabilities are reduced~\cite{Sarff03,Chapman09}.
The self-organized, quasi-single-helicity (QSH) regime that appears spontaneously in high current RFP plasmas also has improved confinement from reduced stochastic magnetic transport~\cite{Lorenzini09}.

An inductive control method called self-similar ramp-down (SSRD) has been shown theoretically to completely stabilize tearing in the RFP~\cite{Nebel02}, but the required inductive programming is yet to be demonstrated in experiments.
In SSRD, both $B_p(t)$ and $B_t(t)$ are ramped down using simultaneous programming of both circuits at a characteristic rate somewhat faster than the plasma's natural $L/R$ time. This inductively sustains a tearing-stable magnetic equilibrium having constant $q(r)$ without the need for dynamo relaxation and its concomitant magnetic fluctuations,
whereas PPCD imparts a large change in $q(r)$.
Further,
advanced inductive control
capable of transitioning between OFCD and SSRD programming on demand and with minimal delay could yield a hybrid, nearly-steady-state scenario that combines the advantages of efficient current sustainment via OFCD and stability control via SSRD using robust inductive current drive~\cite{Sarff08}.

The development of real-time programmability is essential to achieve such advanced inductive control for an RFP plasma. Programmable power supplies are being developed for the MST facility, in which advanced control scenarios can be deployed and tested. Partial power supplies exist for low-current operation, which are used to begin the development of advanced control. Power supplies capable of high-current advanced inductive control experiments in MST are presently under construction. Immediate advantages accrue, since inductive control enables a wide range of new capabilities spanning the breadth of MST's fusion and basic plasma science missions.

In a broader context, active control of MHD instabilities is important for magnetically confined plasmas.  For example, in tokamak experiments, toroidal rotation is believed to have an important effect on MHD stability, where altering the plasma profile and speed can increase the stability of tearing, kink/ballooning, and resistive wall modes
(RWM)~\cite{Gerhardt09, Park13, Sabbagh10, Berkery10, Garofalo02}.
Feedback systems consisting of real-time computation, arrays of magnetic sensors (to measure the toroidal angular plasma momentum) and external actively actuated coils and beam injectors (to drive or drag the rotation) have the potential to maintain plasma stability which is  an important factor in avoiding disruptions in tokamaks~\cite{Goumiri15, Goumiri17}.
Physics-based feedback controllers have been successfully applied, both in the tokamak community with the real-time control schemes based on RAPTOR~\cite{Felici18}, and in the RFP community with the `clean mode control' technique for RMW stabilization in RFX-mod~\cite{Zanca12}.
In the context of the present work, two other prominent examples are classical controllers for $I_p$~\cite{Bettini11} and $F$~\cite{Barp11} in RFX-mod.
Modern control theory approaches have been taken in both RFPs and tokamaks, where advanced feedback control algorithms have been used in RWM control problems for different devices: \cite{Olofsson09,Olofsson13}~for the EXTRAP T2R and RFX-mod RFPs and \cite{Garofalo01, In06, Sabbagh04}~for the DIII-D and NSTX tokamaks are good examples.

%

This paper reports the first systematic tests of real-time control of the programmable power supplies presently available on MST. The initial focus is simultaneous control of the toroidal plasma current, $I_p$, and the dimensionless reversal parameter, $F= B_T(a)(\pi a^2/\Phi)$, where $B_T(a)$ is the toroidal field at the plasma surface, and $\Phi$ is the toroidal flux within the plasma. The reversal parameter tends to track the edge safety factor, $q(a)\propto B_T(a)$, but $\Phi$ is determined primarily by poloidal {\em plasma} current, since the applied toroidal field is small for the RFP. Therefore, $F$ is influenced by the nonlinear relaxation process occurring in the plasma and is not simply proportional to circuit current in a power supply. Controlling $F$ represents a first step toward understanding the influence of nonlinear effects in advanced control of the RFP magnetic equilibrium.

The work begins by building linear models through a system identification method based on experimental data, and then tests these models on independent sets of experimental data. A systematic linear control theory procedure is then used to design a model-based controller which is eventually applied to the MST device control experiment through a fully integrated control software, here dubbed the MST Control System (MCS).
We find that the model-based feedback controllers show some aspects of improved performance relative to simple feed-forward controllers, but the work reveals additional development is required to achieve the desired advanced controllers.

The paper is organized as follows. Section~\ref{sec2} introduces the data-driven models to predict MST reversal parameter $F$ and plasma current $I_\text{p}$ separately then simultaneously.  Section~\ref{sec3} describes the general design of the used controllers. Experimental results of $F$ and $I_\text{p}$ control are shown in Section~\ref{sec4}. Section~\ref{sec5} concludes the paper.

\section{The data-driven modeling}
\label{sec2}

The MST facility~\cite{Dexter91} produces RFP plasmas with current $I_p<0.6$~MA. Its major and minor radii are $R_0=1.5$~m and $a=0.5$~m respectively.
For the work in this paper, low-current plasmas
($I_p\approx 75$~kA, $T_e\approx 60$~eV, density $n_e\approx 1\times 10^{19}$~m$^{-3}$)
were studied using
programmable switching power supplies~\cite{Holly11,Holly15} attached to the poloidal field transformer and toroidal magnet.
These supplies presently have limited current capability, but
upgrades capable of much larger current are under construction. The facility also produces low-current tokamak plasmas.

\begin{figure} 
\centering
\includegraphics [width=1\linewidth]{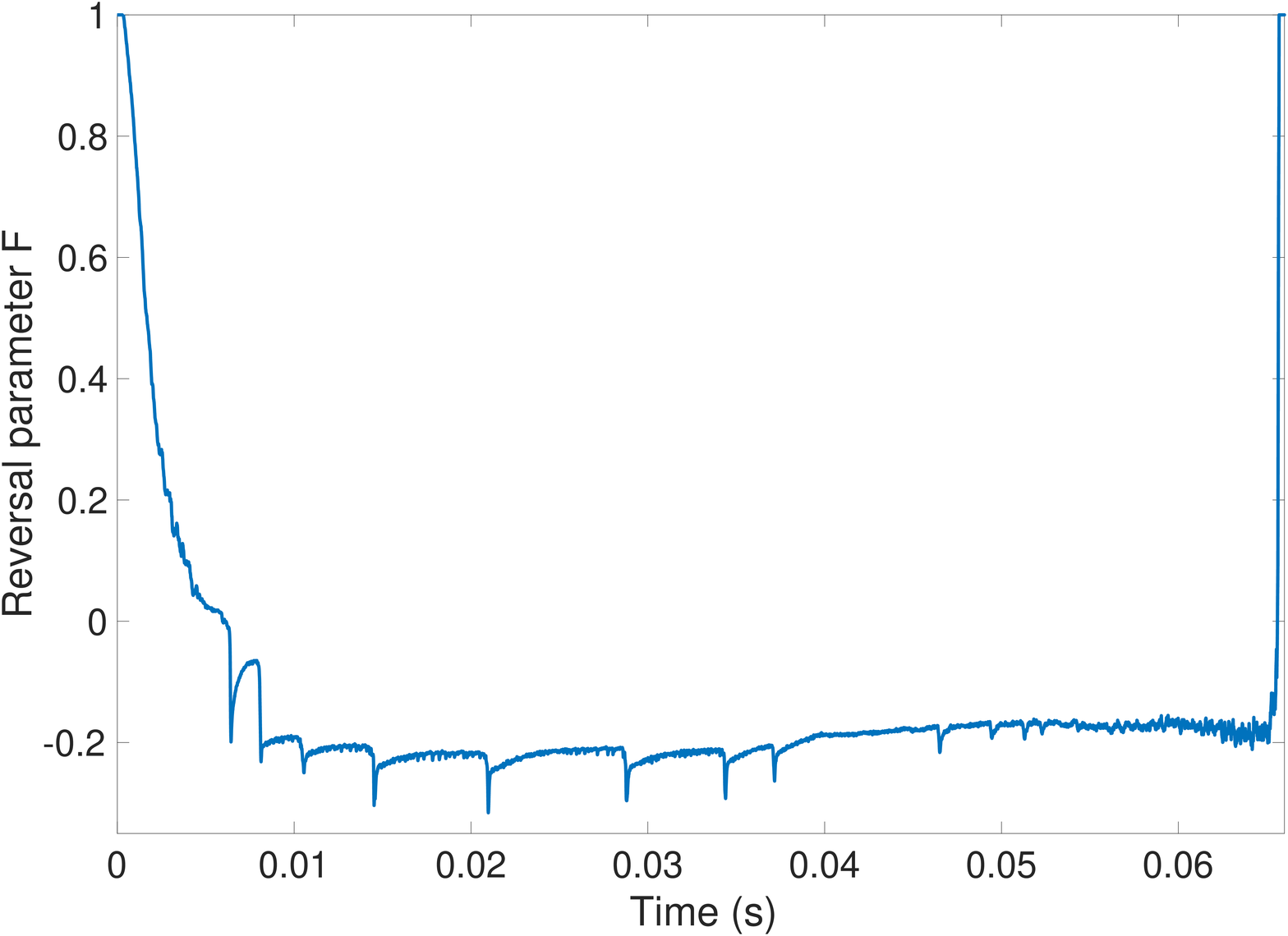}  \hspace{-4.5em}\raisebox{12.7em}{(a)} \\
\includegraphics [width=1\linewidth]{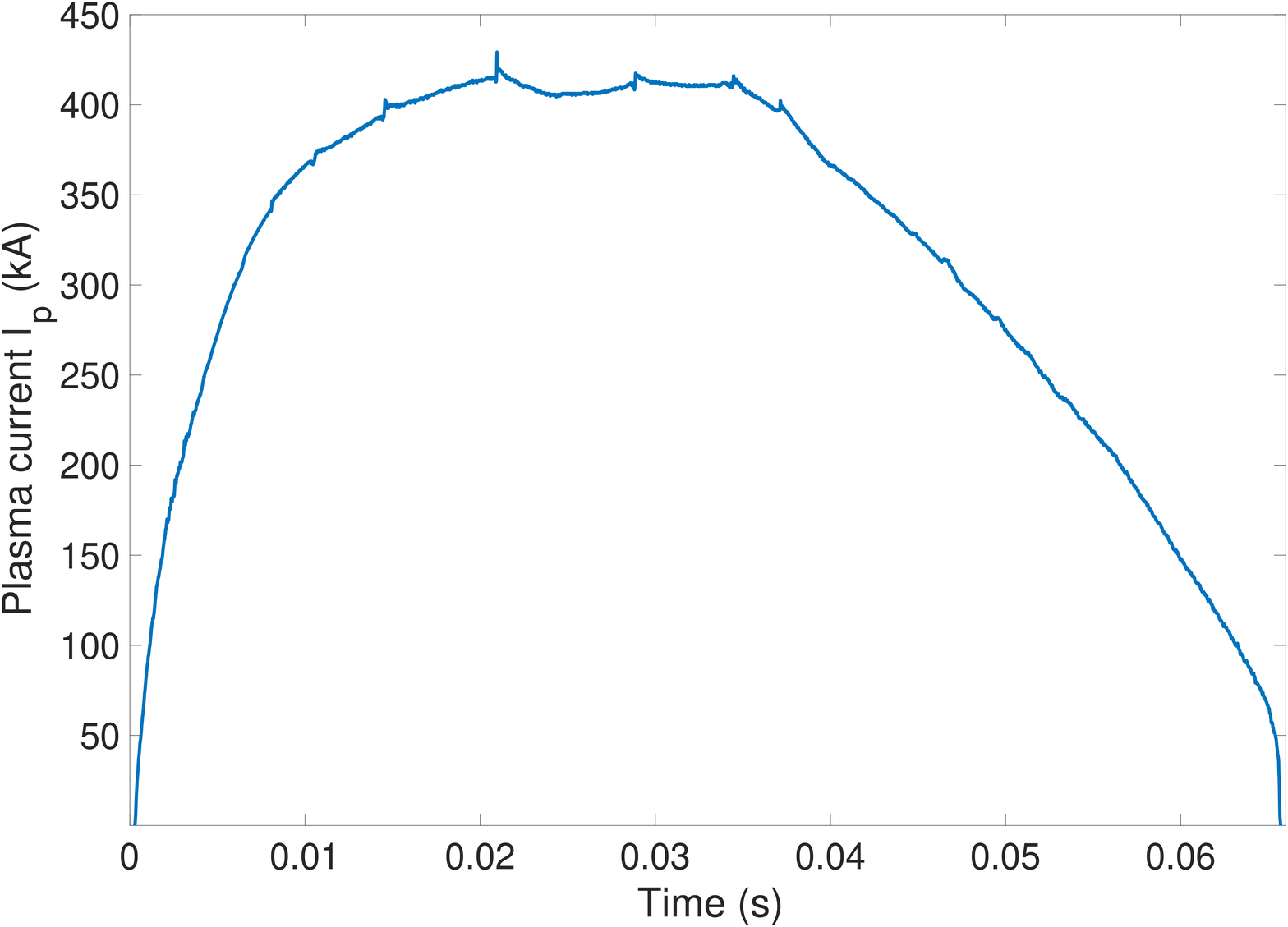}  \hspace{-4.5em}\raisebox{12.7em}{(b)} 
\caption{Experimental data (Shot \# 1161209010) exhibiting sawteeth effect on the dynamics of the reversal parameter $F$ (a) and the plasma current $I_\text{p}$ (b). This shot is a standard MST plasma produced using legacy non-programmable power supplies that attain an approximate ``flattop'' using pulse-forming-network techniques.}
\label{fig1}
\end{figure}

Waveforms for $F(t)$ and $I_p(t)$ in a typical MST discharge are shown in figure~\ref{fig1}. 
The waveforms are punctuated by abrupt quasi-periodic
events, called sawteeth.
These events
result from the magnetic relaxation process related to tearing instabilities that maintain the current profile near marginal stability. Our designed controller is not aiming to control
the fast sawtooth dynamics but rather control the 
slower trend of $F$ and $I_\text{p}$. Therefore sawtooth dynamics will appear and affect our controlled results.

The basic idea of system identification is characterizing a dynamical system from sampled input and output data~\cite{Ljung99}. We assume that our MST device can be approximated by a linear time invariant (LTI) discrete-time system. Convergence results show that it is possible, for a particular class of models, to asymptotically obtain an accurate LTI system description as the size of the sample dataset $N$ and the model order $n$ both goes to infinity, but with $n/N$ vanishing~\cite{Zhu01}.

\begin{figure}
\centering
\begin{center}
\tikzstyle{block} = [draw, fill=yellow!10, rectangle, minimum height=2em, minimum width=4em]
\tikzstyle{blockk} = [draw, fill=yellow!10, rectangle, minimum height=4em, minimum width=4em]
\tikzstyle{blockw} = [draw, fill=green!10, rectangle, minimum height=2em, minimum width=4em]
\tikzstyle{sum} = [draw, fill=blue!20, circle, node distance=0.9cm]
\tikzstyle{input} = [coordinate]
\tikzstyle{output} = [coordinate]
\tikzstyle{pinstyle} = [pin edge={to-,thin,black}]

\begin{tikzpicture}[auto, node distance=2cm,>=latex']
    \node (r) [fill,circle,inner sep = 0pt,minimum size=1pt] {};
    \node [blockw, right=0.5 of r] (MIMO) {MIMO controller};
    \node [block, right of=MIMO ,node distance=4.1cm] (MST) {MST};
    \draw [connector] (MIMO.8)  -- ++ (2,0)  (MST)  node[pos=0.5, above] {$I_{tg}$};
    \draw [connector] (MIMO.-8) -- ++ (2,0)  (MST)  node[pos=0.5, below] {$I_{pg}$};
     \node [output, right of=MST] (output) {};
    \draw [connector] (MST.15)   -- ++ (1,0)   node[pos=0.6, above, yshift = 2] {$F$} -| ++ (0,-1.8) -| node[pos=0.2, above, yshift = -1] {} (r) to (MIMO);
   \draw [connector] (MST.-15)   -- ++ (0.5,0)   node[pos=0.6, below, yshift = -2] {$I_p$} -| ++ (0,-1.1) -|  node[pos=0.2, below, yshift = -1] {}  (MIMO.270);
\end{tikzpicture}

\begin{tikzpicture}[auto, node distance=2cm,>=latex']
    \node (r) [fill,circle,inner sep=0pt,minimum size=1pt] {};
    \node [blockw, right=0.5 of r] (SISO1) {SISO controller};
    \node[blockw, below=0.25 of SISO1] (SISO2) {SISO controller};
    \node [blockk, right=2 of SISO1, yshift=-13] (MST) {MST};
    \draw [connector] (SISO1.0.3)  -- ++ (2,0)  (MST)  node[pos=0.5, above] {$I_{tg}$};
    \draw [connector] (SISO2.-0.3) -- ++ (2,0)  (MST)  node[pos=0.5, below] {$I_{pg}$};
     \node [output, right of=MST] (output) {};
    \draw [connector] (MST.15)   -- ++ (1,0)   node[pos=0.5, above, yshift = 2] {$F$} -| ++ (0,-1.8)  -|  (r) to (SISO1);
   \draw [connector] (MST.-15)   -- ++ (0.5,0)   node[pos=0.6, below, yshift = -2] {$I_p$} -| ++ (0,-1.1) -|  (SISO2.290);
\end{tikzpicture}
\caption{Schematic of the MIMO and the two-loop control design approaches.}
\label{Schematic}
\end{center}
\end{figure}
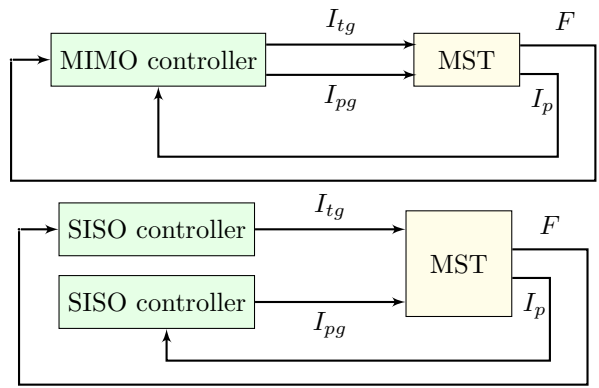

Three models are constructed here:  A first single input single output (SISO) model of the reversal parameter $F$, a second (SISO) model of the plasma current $I_\text{p}$ and finally a multiple inputs multiple outputs (MIMO) model for the coupled dynamics of $F$ and $I_\text{p}$. Each model is going to have a corresponding controller and be applied in real time experimentally to the MST device. The schemes are depicted in Figure~\ref{Schematic}. The procedures are detailed in the following subsections.

\subsection{ The $F$ model}
\label{sec2a}

\begin{figure} 
\centering
\includegraphics [width=1\linewidth]{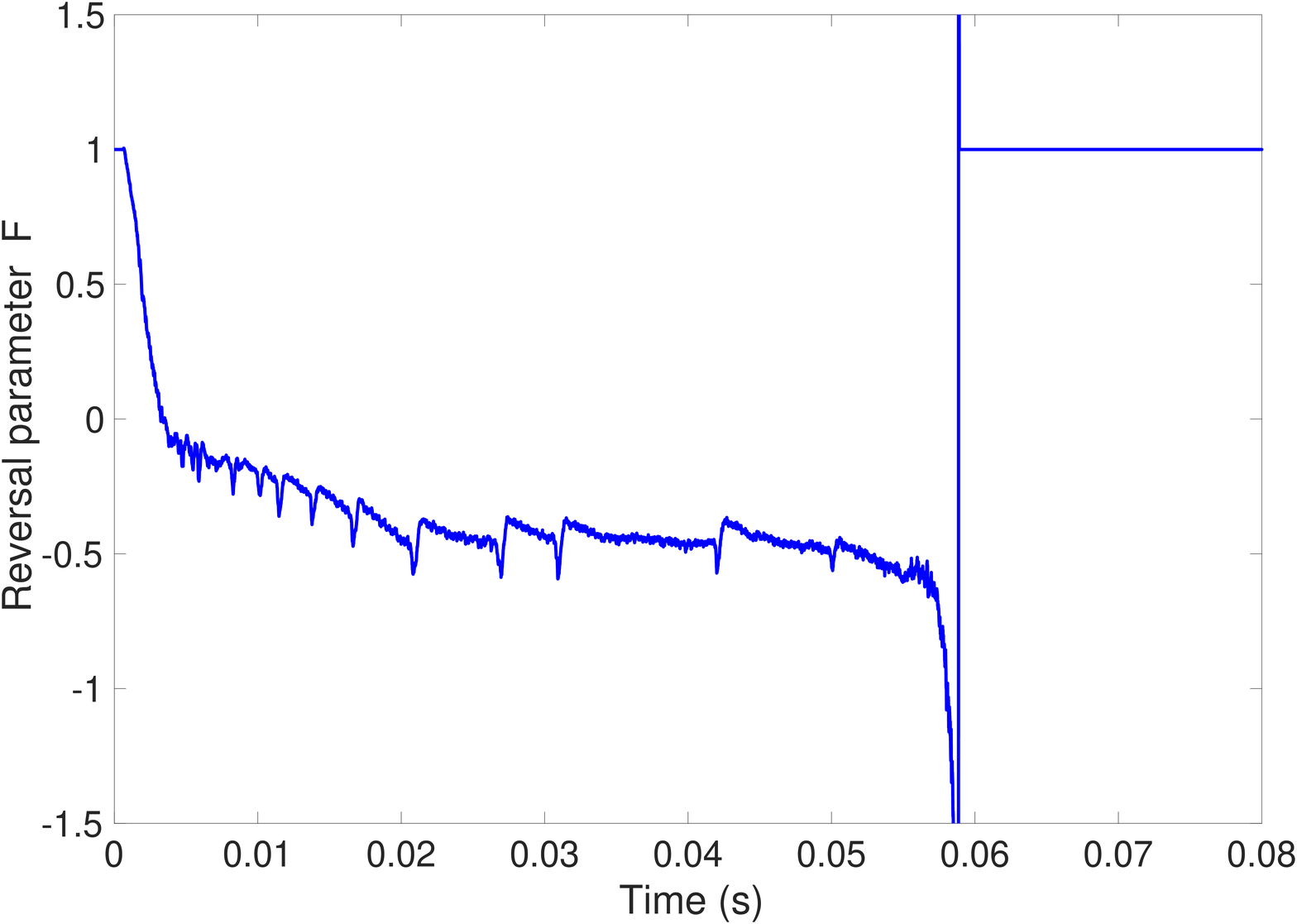}  \hspace{-5em}\raisebox{13.5em}{(a)} \\
\includegraphics [width=1\linewidth]{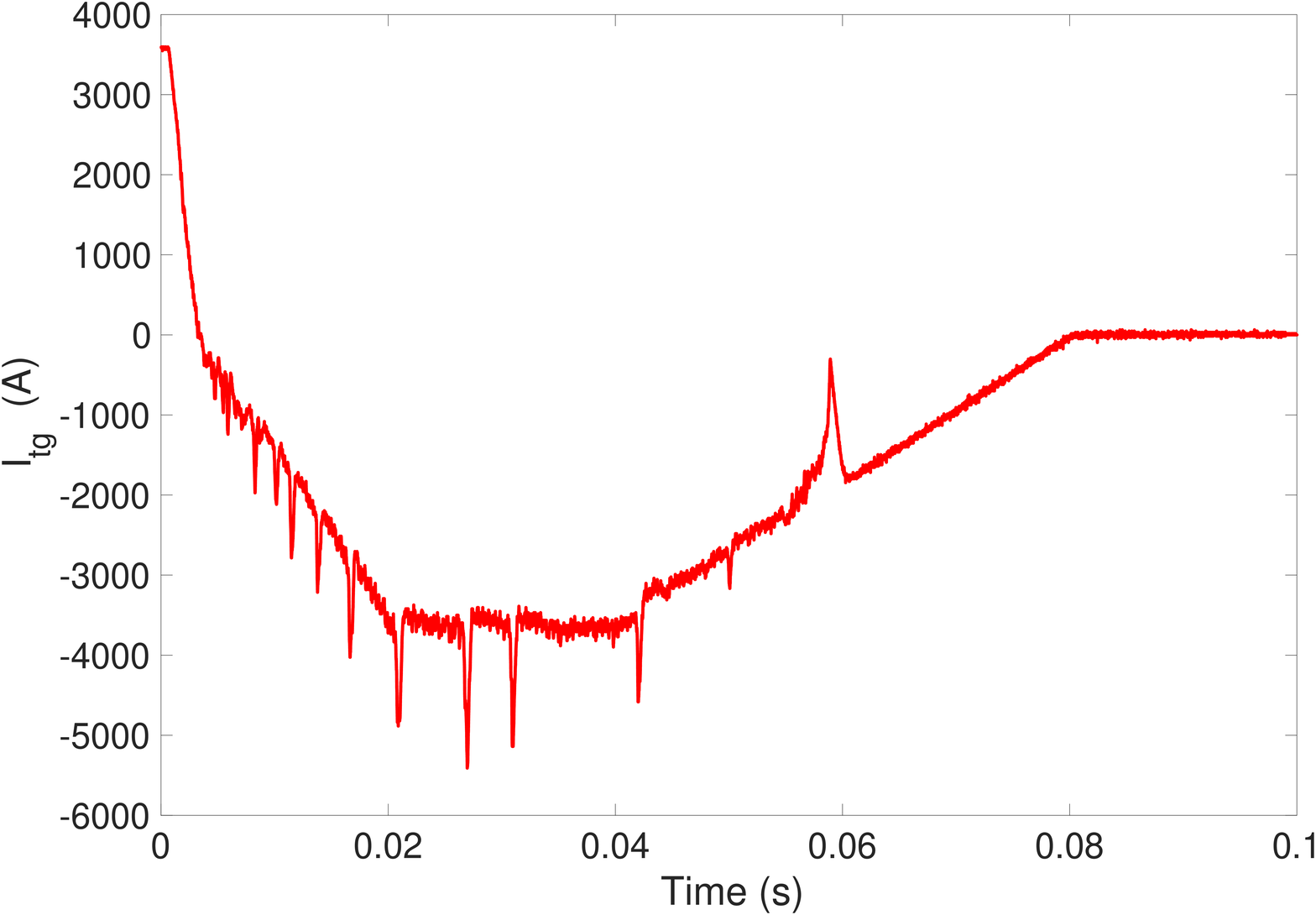}  \hspace{-5em}\raisebox{13.5em}{(b)} 
\caption{ An experimental data (Shot \# 1170822156) showing the relationship between the input ($I_{tg}$) and the output ($F$) of our reversal parameter model.}
\label{figg1}
\end{figure}

The collected data come from a hundred shots (estimation data set) where the input data (actuator) is the primary current $I_{tg}$ of the $B_\text{t}$ programmable power supply and the output data (sensor measurement) is the reversal parameter $F$ ranging from $0$ to $-0.4$. Figure~\ref{figg1} represents one example of the data collected. we can see that up to the end of the plateau period ($t=0.04$~s), the dynamic shows a linear behavior. This will be the focus of our modeling.

A system identification process was used to develop a linear state-space response model to the system. This model will then be used to design an optimal control law. 
The discrete linear state-space response model can be written of the form
\begin{equation}
\begin{split}
 {x}_{k+1} &= A x_k + B u_k , \label{ss1}\\
 y_k &= C x_k,
\end{split}
\end{equation}
where the physical actuator value $u_k=I_{tg}$ is the primary current at a certain iteration and the measurement of the system output $y= F$ is the corresponding reversal parameter. $A \in \mathbb{R}^{\, (n_x) \times (n_x)}$, $B \in \mathbb{R}^{\,(n_x) \times 1}$, and $C \in \mathbb{R}^{\,1 \times (n_x)}$ which respectively are called the dynamics, control and sensor matrices, identified through system identification.

The subspace method~\cite{Ljung99} for state-space model identification, part of the Matlab System Identification Toolbox, was used to find the optimal system matrices for a prescribed number of states $n_x$ (model-order) that best fitted the estimation data set. The optimal choice of model-order was then found by identifying a set of models for a small range of $n_x$, simulating the identified models using the inputs from the validation dataset (a different set of shots), and comparing how well each model predicted the output of the validation dataset. Models with too low number of states fail to capture the main dynamics of the system, while models with excessive number of states overfit the noise in the estimation data set, degrading prediction of the validation dataset. 
 
\begin{figure} 
\centering
\includegraphics [width=1\linewidth]{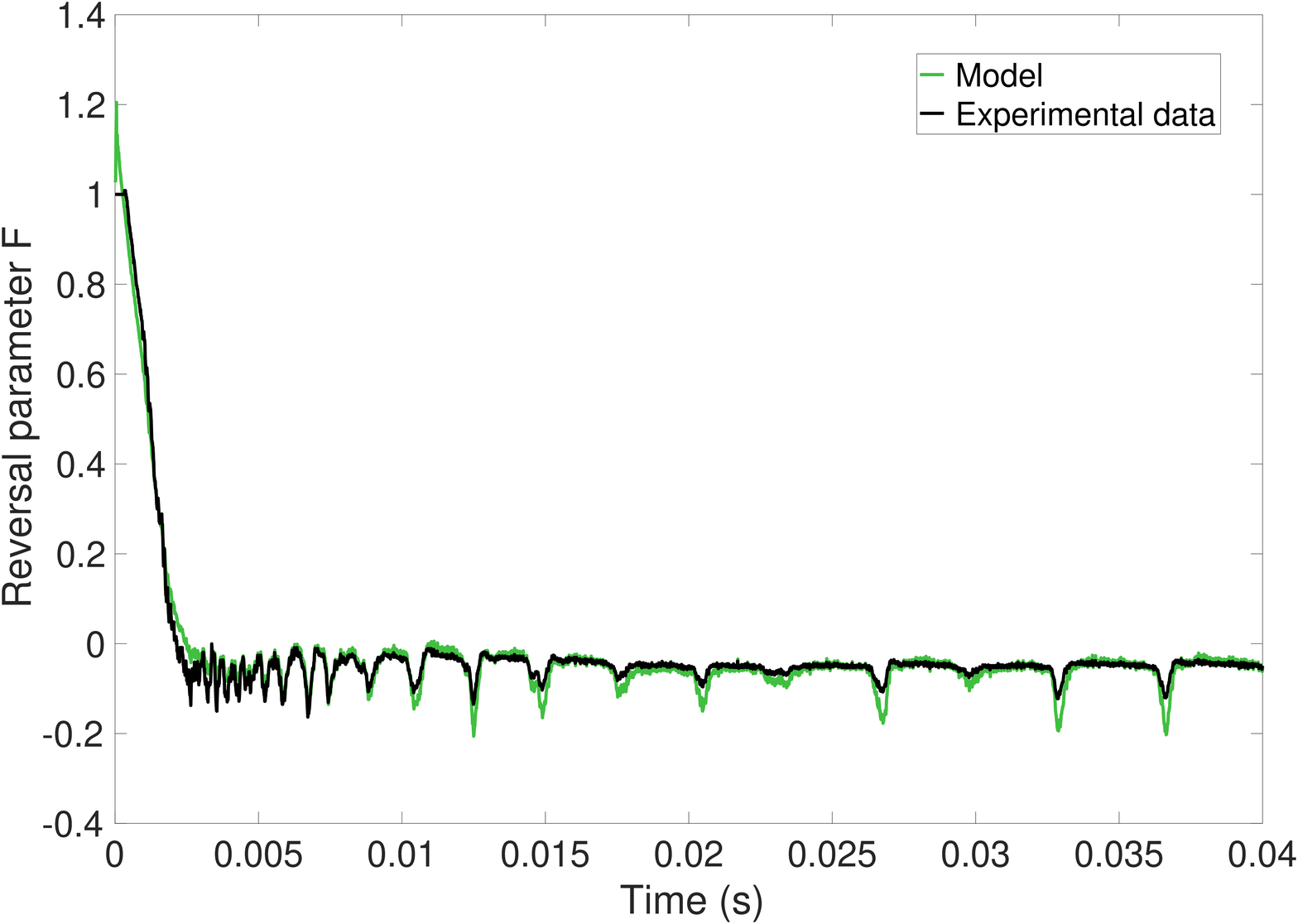} 
\caption{Comparison of output ($F$) predicted by the identified model to the actual reversal parameter $F$ of the validation data.}
\label{figg2}
\end{figure}
A comparison of the outputs of the optimal model, which was found to be of the order of three, to the validation data is shown in figure~\ref{figg2}, showing good agreement in $F$ (root mean square error of 15\%).
The benchmarking of the model against several real data shots is a necessary first step as this model will be used for our control design testing. 
An exact plasma model is not a major concern as feedback control can be performed to tolerate errors in the model. The key is to ensure that the $F$ model does not deviate drastically from the actual time evolution in order to prevent control system instabilities from dominating plasma physics dynamics.
\subsection{ The $I_p$ model}
\label{sec2c}
\begin{figure} 
\centering
\includegraphics [width=1\linewidth]{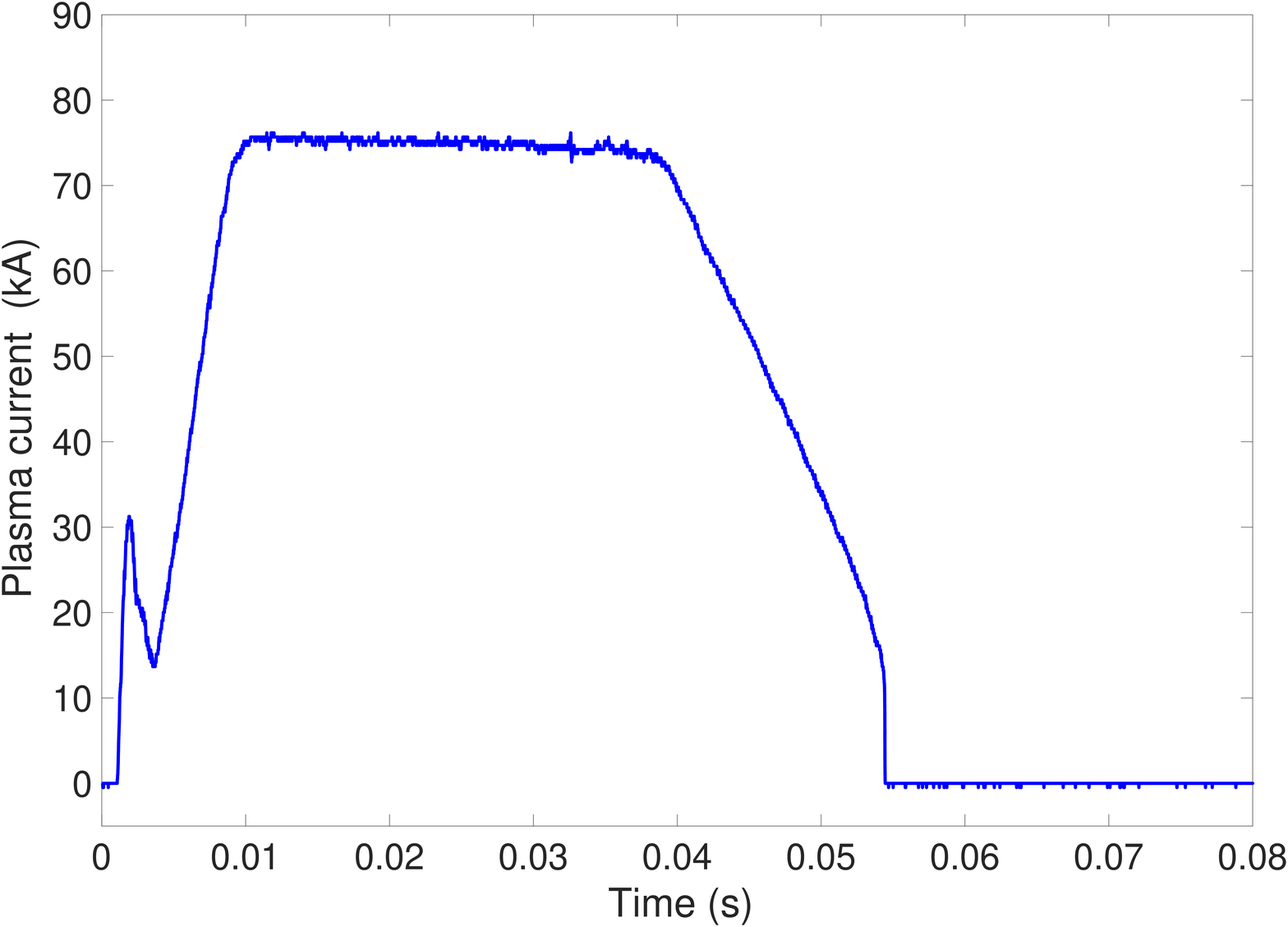}  \hspace{-5em}\raisebox{13.5em}{(a)} \\
\includegraphics [width=1\linewidth]{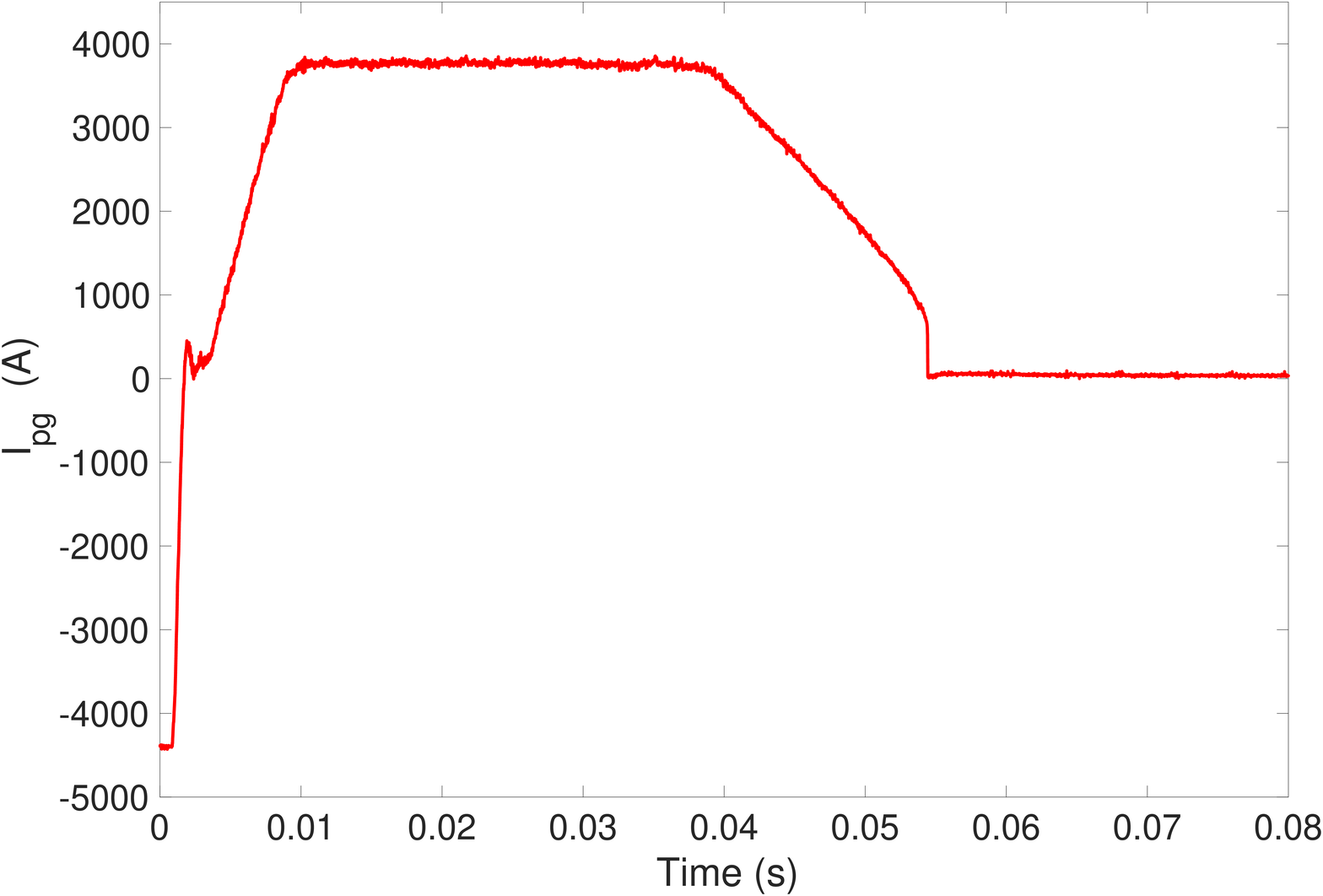}  \hspace{-5em}\raisebox{13.5em}{(b)} 
\caption{ An experimental data (Shot \# 1180207047) showing the relationship between the input ($I_{pg}$) and the output ($I_\text{p}$) of our plasma current model.}
\label{figg3}
\end{figure}

The dedicated data come from about 50 shots where the input data (actuator) is the primary current $I_{pg}$ of the $B_\text{p}$ programmable power supply and the output data (sensor measurement) is the plasma current $I_\text{p}$. Figure~\ref{figg3} represents an example of the data collected. We can see that during this shot period, the dynamics shows a linear behavior but with a slight decay of plasma current around $t=0.035$~s that persists until the end of the shot. This phenomenon is due to plasma resistance. Our model does not intend to capture this current decrease as it is assumed linear.

\begin{figure} 
\centering
\includegraphics [width=1\linewidth]{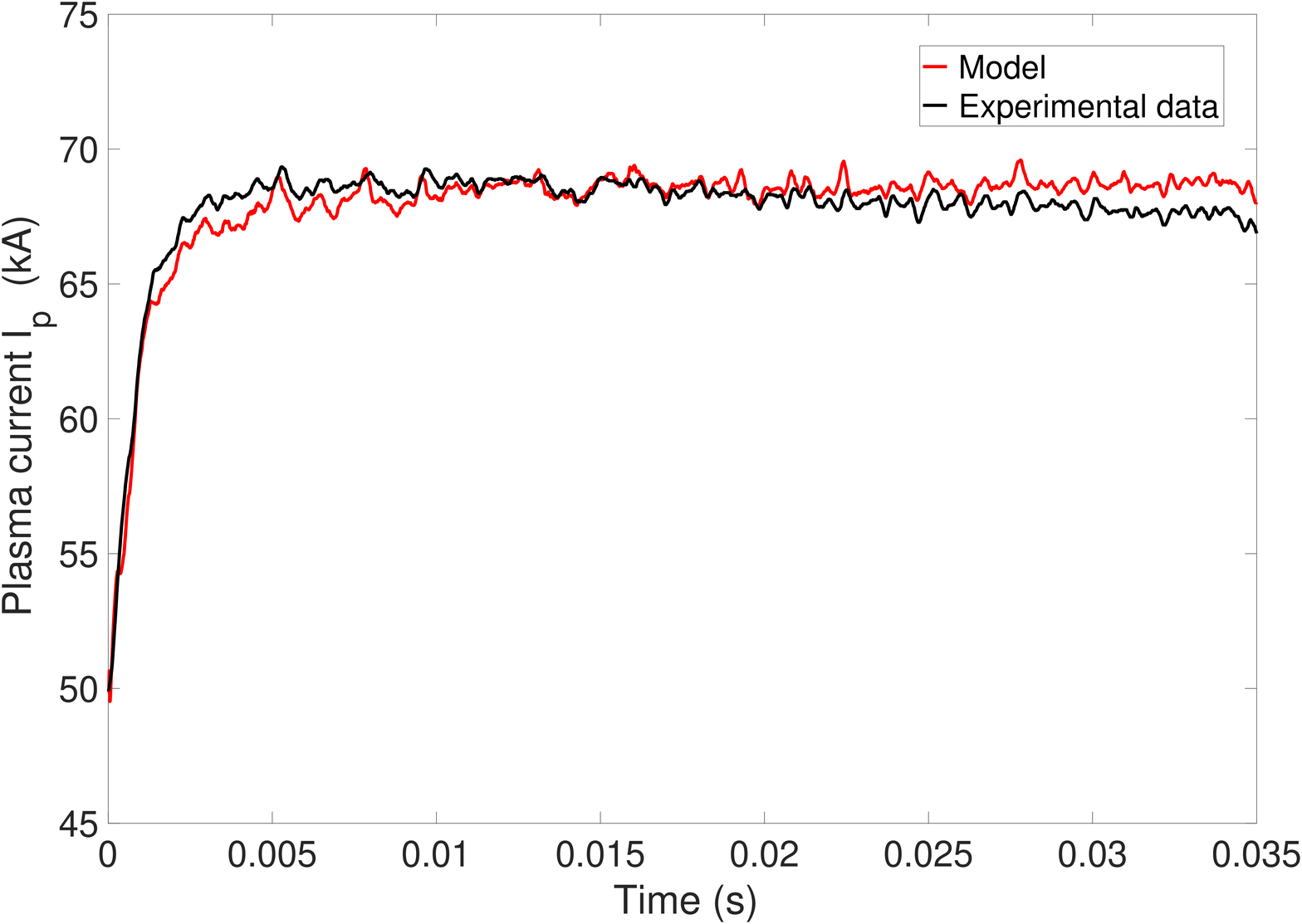} 
\caption{ A comparisons of output ($I_\text{p}$) predicted by the identified model to the actual plasma current of the validation data}
\label{figg4} 
\end{figure}

A comparison of the outputs of the optimal model to the validation data is shown in figure~\ref{figg4}, showing good agreement in $I_\text{p}$ up to the time where the plasma resistance effect starts to become noticeable ($t=0.035$~s). The model is linear, so for a constant input ($I_{pg}$), it predicts a steady plasma current whereas the experiments show a little decrease towards the end due to the plasma resistance.

\subsection{ The coupled $F$ and $I_\text{p}$ model}
\label{sec2b}
During the modeling, we started by a SISO model for $F$ and $I_\text{p}$, as the purpose was to test and control each variable independently on MST. Naturally we split the system into two single-input-single-output loops, and we use the models of $F$ and $I_\text{p}$ described above and its controllers to operate it. Once these types of controllers work on MST, we use the fact that these two entities are dynamically coupled to build a coupled MIMO model that has the two primary currents $I_{tg}$ and $I_{pg}$ as inputs and both $F$ and $I_\text{p}$ as outputs.

\begin{figure} 
\centering
\includegraphics [width=1\linewidth]{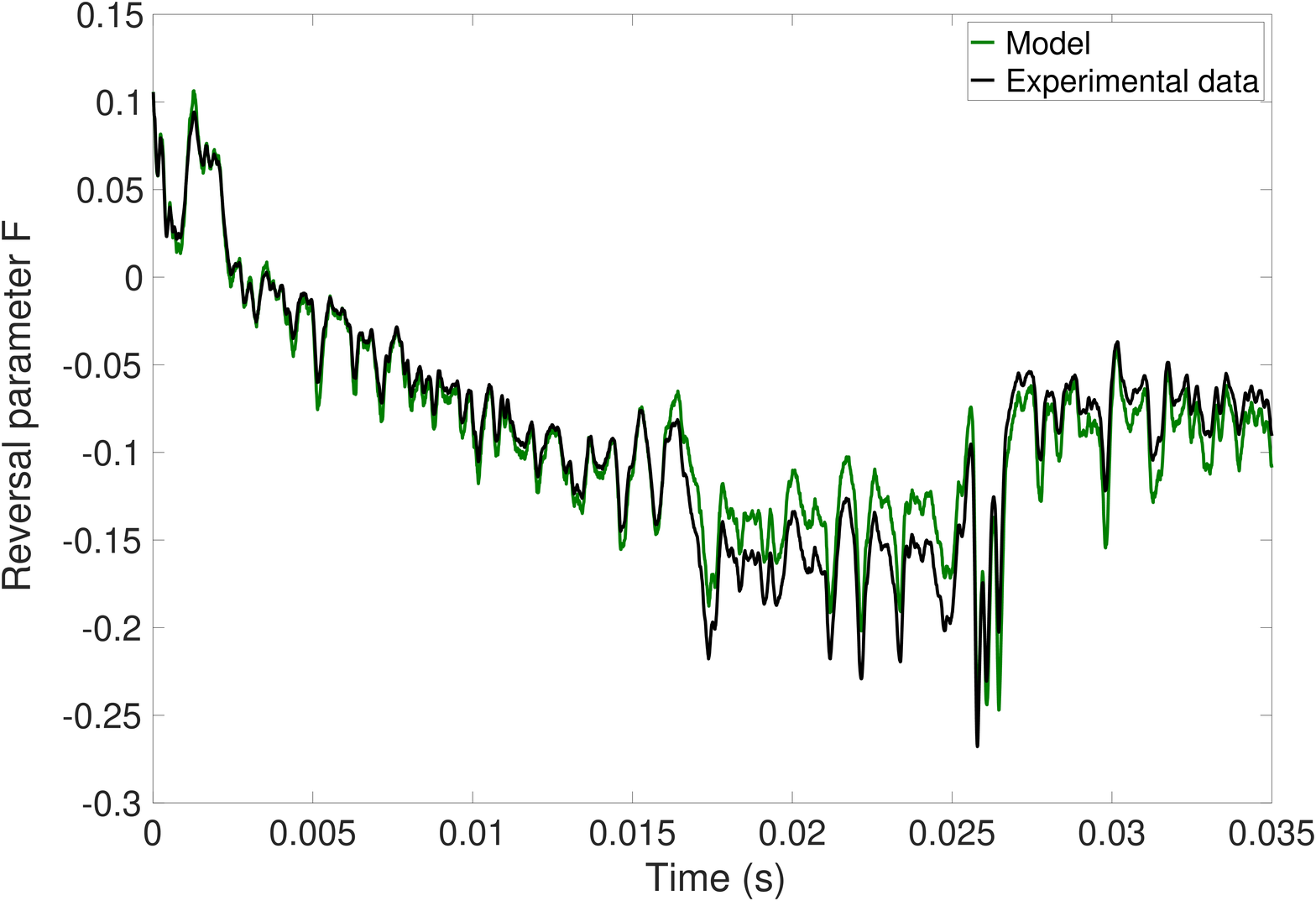}  \hspace{-5em}\raisebox{11.5em}{(a)} \\
\includegraphics [width=1\linewidth]{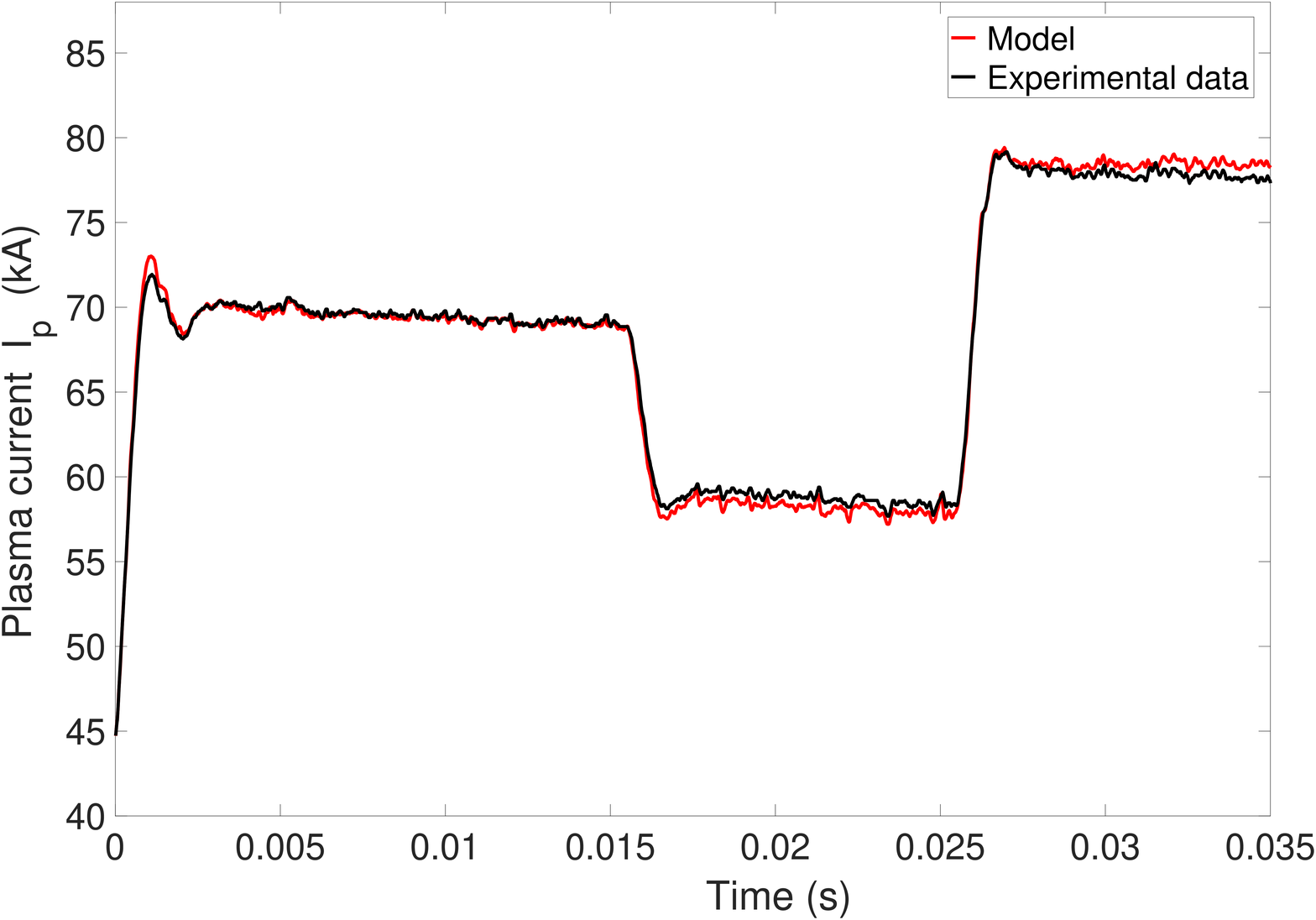}  \hspace{-6em}\raisebox{10.5em}{(b)} 
\caption{ A comparisons of outputs ($F$ (a) and $I_\text{p}$ (b)) predicted by the identified model to the actual reversal parameter and plasma current of the validation data}
\label{figg5}
\end{figure}

A comparison of the outputs of the optimal model to the validation data is shown in figure~\ref{figg5}, showing good agreement in $F$ and $I_\text{p}$. We can notice that the $F$ prediction of the coupled model loses some precision (lower fitting) compared to its prediction in the individual $F$ model. This is due to the coupling consideration. It will produce though a better controller design due to the additional dynamics information it encapsulates (if we compare it to the split system).

\section{ Control design}
\label{sec3}

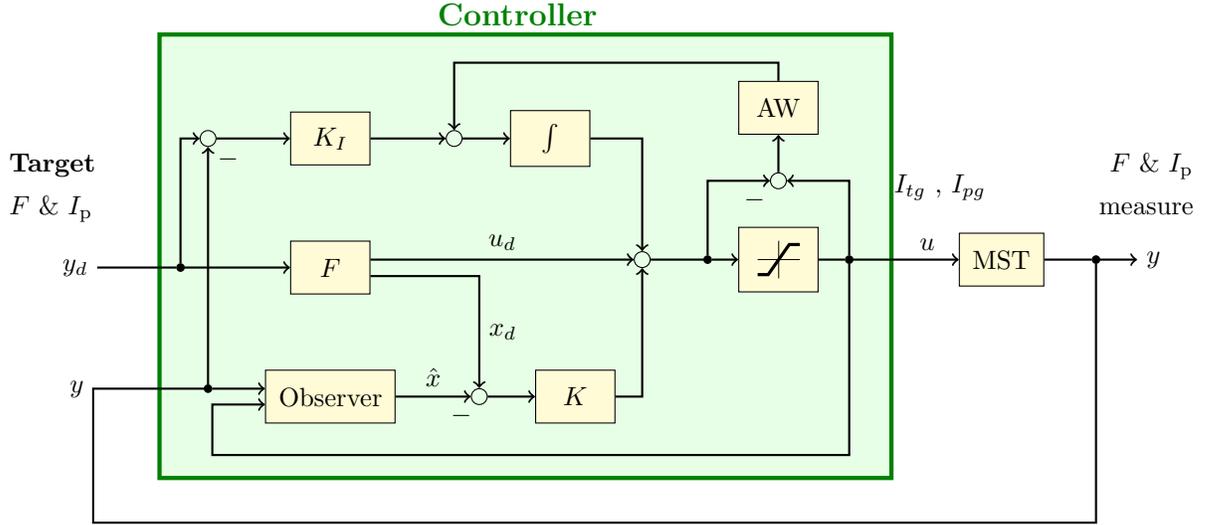
\begin{figure*}
\begin{tikzpicture}[x=0.7cm]
		\draw[green!50!black, fill=green!10, ultra thick] (1.6,-2.8) rectangle (15.5,3.1);
		\node[green!50!black, anchor=south, font=\large\bfseries] at (8.4,3.1) {Controller};

		\node (r) {$y_{d}$};
		\node[above=0.25 of r, text width=5em] {\baselineskip=16pt \textbf{Target} \\$F$ \& $I_\text{p}$ \par};
		\node[junction, right=1.5 of r] (r in) {};
		\node[block, right=2 of r in] (F) {$F$};
		\node[block, below=of F] (L) {Observer};
		\node[sum, right=of L] (sum feedback) {};
		\node[block, right=0.9 of sum feedback] (K) {$K$};
		\node[sum, right=5 of F, yshift=3] (sum inputs) {};
		\node[junction, right=1 of sum inputs] (before sat) {};
		\node[block, right=0.5 of before sat] (sat) {
			\begin{tikzpicture}
				\draw[very thin] (-.4,0) -- (.4,0) (0,-.25) -- (0,.25);
				\draw[very thick] (-.4,-.2) -- (-.2,-.2) -- (.2,.2) -- (.4,.2);
			\end{tikzpicture}
		};
		\node[junction, right=0.5 of sat] (after sat) {};
		\node[block, right=2 of after sat] (P) {MST};
		\node[below=0.15 of P, xshift=-0.5ex, text width=4em, align=center] {\baselineskip=16pt \par};
		\node[junction, right=0.9 of P] (P out) {};
		\node[right=0.7 of P out] (y) {$y$};
		\node[above=0.25 of y, xshift=-1em, text width=5em, align=right] {\baselineskip=16pt $F$ \& $I_\text{p}$ measure \par};
		\node[junction, left=1 of L, yshift=3] (y in) {};
		\node[coord, left=1 of L, yshift=-3] (sub y in) {};
		\node[coord, label=left:$y$, left=2.1 of y in] (y input) {};
		\node[block, above=of F] (Ki) {$K_I$};
		\node[sum] (sum lqi) at (Ki -| y in) {};
		\node[sum, right=of Ki] (AW out) {};
		\node[block, right=0.9 of AW out] (integrator) {$\int$};
		\node[sum, above=0.5 of sat] (sum AW) {};
		\node[block, above=0.5 of sum AW] (AW) {AW};
		
		\draw[connector] (r) to (r in) to (F);
		\draw[connector] (F.east |- sum inputs) to node [above] {$u_{d}$} (sum inputs);
		\draw[connector] (F)[yshift=-12] -| node [right, near end] {$x_{d}$} (sum feedback);
		\draw[connector] (sum feedback) to (K);
		\draw[connector] (K) -| (sum inputs);
		\draw[connector] (sum inputs) to (before sat) to (sat);
		\draw[connector] (sat) to (after sat) to ++(down:2.6) -| (sub y in) to (sub y in -| L.west);
		\draw[connector] (after sat) to node [above, pos=0.7] (u) {$u$} (P);
		\node[above=0.25 of u, text width=3.5em, xshift=0.5em] {\baselineskip=16pt $I_{tg}$ , $I_{pg}$ \par};
		\draw[connector] (P) to (P out) to (y);
		\draw[connector] (P out) -- ++(down:3.5) -| (y input) to (y in) to (y in -| L.west);
		\draw[connector] (L) to node [above] {$\hat x$} node [below, very near end] {$-$} (sum feedback);
		\draw[connector] (r in) |- (sum lqi);
		\draw[connector] (y in) to node [right, pos=0.95] {$-$} (sum lqi);
		\draw[connector] (sum lqi) to (Ki);
		\draw[connector] (Ki) to (AW out);
		\draw[connector] (AW out) to (integrator);
		\draw[connector] (integrator) -| (sum inputs);
		\draw[connector] (before sat) |- node [below, very near end] {$-$} (sum AW);
		\draw[connector] (after sat) |- (sum AW);
		\draw[connector] (sum AW) to (AW);
		\draw[connector] (AW) to ++(up:0.6) -| (AW out);
	\end{tikzpicture}

\caption{Global schematic of the controller that combine a feedforward $(F)$, a LQR $(K)$, an observer, an integrator $(K_I)$ and an anti-windup (AW).}
\label{fig:model1}
\end{figure*}

Once the identified models are built, we use them to design Linear Quadratic Gaussian (LQG) controllers. This type of controllers minimizes a cost function of the form~\cite{SandP, Lewis, AandM}
\begin{align}
J = \int_0^T { x(t)^\mathsf{T}\! Q x(t) + u(t)^\mathsf{T}\! R u(t) + x_i(t)^\mathsf{T}\! Q_i x_i(t) \, dt},
\end{align}
where $x(t)$ is the internal state of the system at time $t$, $u(t)$ is the control input, $x_i(t)$ is the integral of the tracking error (between the targeted and actual values). The controller optimizes the use of actuators according to the weights in $Q$ and $R$, which are free design parameters, and also ensures reference tracking with the integral action tailored by choice of the free design parameters in $Q_i$. A Kalman filter~\cite{SandP, Lewis, AandM} is embedded in the resulting control law, which optimally estimates the unmeasured states $x(t)$ based on the measurements $y(t)$, taking into account the process and measurement noise levels. An anti-windup scheme is implemented to keep the actuator requests from winding up far beyond their saturation levels by feeding back a signal proportional to an integral of the unrealized actuation. Figure \ref{fig:model1} represents the schematic of this controller design. 


The same control design will be used for both One-Input One-output models that control $F$ using $I_{tg}$ and $I_\text{p}$ using $I_{pg}$, and the Two-input Two-output model (MIMO) that controls simultaneously $F$ and $I_\text{p}$ using primary currents $I_{tg}$ and $I_{pg}$. The only difference will be the dimension of the model inputs and outputs which change from a single variable to a vector variable; the controller dimensions will adapt accordingly. More details about the control theory and design can be found in~\cite{Astrom10, Skogestad05} but will be summarized succinctly in this section. As shown in Figure \ref{fig:model1}, the controller design has five main components:
\subsection{Feedforward design $F$} 
The purpose here is to force the plasma current $I_p$ and the revesal parameter $F$ to reach a target state $x_d$ such that the sensor output $y$ matches a reference signal~$y_d$. In the final implementation, all one should have to prescribe is $y_d$ (e.g., the desired plasma current value $I_p$ and the desired reversal parameter $F$). The target state $x_d$ and the corresponding input $u_d$ are found by solving equations~(\ref{ss1}) at steady state:
\begin{equation} \label{ss2}
\begin{split}
 0 &= A x_d + B u_d, \\
 y_d &= C x_d.
 \end{split}
\end{equation}
We then solve for $x_d$ and $u_d$ by writing (\ref{ss2}) in matrix form
\begin{equation}
\left(\! \begin{array}{c}  x_{d} \\ u_{d}\end{array}\!\right)
  ={ \left(\! \begin{array}{cc} A  & B \\ C & 0 \end{array} \! \right)}^{-1} \left(\! \begin{array}{c} 0 \\ I    \end{array}  \!\right) y_{d} = \left(\! \begin{array}{c} F_x \\ F_u    \end{array}  \!\right) y_{d}.
\label{steadystate}
\end{equation}
Once the desired target states $\left( x_{d} , u_{d} \right)$ are established, the controller is designed based on the model then tested on the MST device to determine if the controller can track and reach the desired $F$ and $I_p$ values in the vicinity of the equilibrium.

If the model of the dynamics has no errors or uncertainties (which is never the case) and is stable, a feedforward controller is enough to reach the target.  $ F_u$ and $ F_x $ are the feedforward gains corresponding  to the input and state respectively.
The total feedforward gain~$F$  depends on the matrices $A$, $B$, $C$ and $K$ (explained in the following subsection).
    
\subsection{Linear quadratic regulator (LQR) design $K$} 

The feedback control law links the input $u$ to the state $x$ by
\begin{equation}
   u = u_{d} - K(x - x_{d}) = - Kx + Fy_{d},
   \label{eqn:ctrllaw_ff}
\end{equation}
where $K$ is the feedback control gain to be determined from control design and $F = F_u + K F_x$ is the total feedforward gain.  Therefore, the resulting closed-loop system of equations~(\ref{ss1}) can be written as
\begin{equation}
\begin{aligned}
      \dot{x} &= (A-BK) x + BF y_{d}, \\
      y &= C x.
\end{aligned}\label{eq:4}
\end{equation}

A  standard linear control technique (linear-quadratic regulators)~\cite{SandP, AandM} is used in order to determine the gains $K$ while minimizing a quadratic cost function.

\subsection{Observer design} 

The feedback law~\eqref{eqn:ctrllaw_ff} requires the knowledge of the full state~$x$.  However, in our actual system-identified model, we don't know the state; we don't even know what the state represents, we only measure the inputs-outputs. However, we may reconstruct an estimate of the state from the available sensor measurements using an {\em observer}.
The observer will then reconstruct the state estimate~$\hat x$, with dynamics given by
\begin{equation}
		\dot{\hat{x}} =  A \hat{x} + B u + L (y - C \hat{x}) 
			= (A- L C) \hat{x} + B u + L y,
		\label{obs}
\end{equation}
where the matrices $A,B$ and~$C$ are the same as those in the model~(\ref{ss1}), and $L$ is a matrix of gains chosen such that the state estimate converges quickly relative to the system's dynamics.
Using our linear model, we design an optimal observer (Kalman filter)\cite{SandP, AandM} to find~$L$.

The observer generates an estimate of the state from the physics model as represented by the state matrix, the inputs and outputs, and once combined to the feedback controller, it forms a \text{linear quadratic Gaussian compensator}~\cite{SandP, AandM}.

\subsection{Integrator design $K_I$} 

The goal is to track both the desired plasma current and reversal parameter values (reference tracking). In order to do that, the steady state error between the output (measured) and the target profile has to be minimized by using an integrator and introducing a new state variable~$z$ that is the integral of the error:
\begin{equation}
	\dot{z} = y_{d} - y = y_{d} - C x.
	\label{integral}
\end{equation}
The new feedback law can be then written as
\begin{equation}
  u  = u_d + K (x_d - x) + K_I \!\!\int (y_d - y)
\end{equation}
where $K_I$ be the gain of the integrator.

\subsection{Anti-windup design $AW$}
A drawback of integral control is that if the actuator values are limited to some range as in our case, then the integrator can accumulate error when the actuator is ``saturated,'' resulting in poor transient performance, a phenomenon known as ``integrator windup.'' 

We use a standard anti-windup scheme~\cite{AandM, Lewis} in which one feeds back the difference between the desired value of~$u$ and its actual (possibly saturated) value to eliminate this effect.


\section{ Experimental results}
\label{sec4}
\subsection{Hardware and software setup}

%


In PPS operation on MST, a real-time Linux host for the MST Control System (MCS) provides demand waveforms clocked to a 10 kHz sampling rate via a 16-bit digital-to-analog converter (D-tAcq ACQ196). The analog output voltages are fed to one or both of the two programmable power supplies $B_\text{t}$-PPS and  $B_\text{p}$-PPS. These supplies are IGBT-based switching supplies with bipolar outputs and a switching frequency of $10 \text{kHZ}$~\cite{Holly11, Holly15}. Each supply is powered by its own capacitor bank. The supplies have small-signal bandwidths of a few $\text{kHZ}$ or less. Each supply uses local feedback to provide an output current proportional to the demand voltage from the controller. The supplies have current limiting to clamp their outputs at a safe level independent of the demand voltage.
The $B_\text{t}$-PPS drives the primary of a 40:1 transformer whose secondary is the aluminum vacuum vessel, generating the toroidal field Bt. The  $B_\text{t}$-PPS is capable of an output current up to +/- 25 kA at a voltage up to +/- 1800 V.
The  $B_\text{p}$-PPS drives the primary of a 20:1 transformer whose secondary is the plasma, generating the plasma current $I_p$.  The experiments described here use a prototype version of the BP PPS which has a limited output capability of +/- 4.8 kA at a voltage up to +/- 2700 V.
Although they are referred to as programmable power supplies, the supplies themselves are not pre-programmed, but act as voltage-controlled, current-output amplifiers to produce the current demanded by their input voltages. The simplest control method uses the MCS to generate a preprogrammed PPS input voltage waveform which yields the desired PPS output current.

Although this is closed-loop control with respect to the supply itself, since it is open-loop with respect to the transformer and plasma impedances, we refer to it as 'open-loop' (feedforward) control, labeled 'MST FF data' in the results presented below. By contrast, the more comprehensive 'closed-loop' (feedback) control is implemented ('MST CL data') to respond in real time to changes resulting from transformer and plasma impedances. During each clock cycle in such closed-loop experiments, MST data is input to the MCS software, processed, and the real-time demand value output to the PPS digitizers before the next clock cycle.

For example, while in open-loop control, one only approximately controls $B_\text{t}(a)$, in closed-loop control, one is able to control the field reversal parameter $F = B_\text{t} (a)/<B_\text{t}>$ after startup. The output current is automatically adjusted in real time to do so.
A demonstration of open-loop control will be superposed to the closed loop feedback control when the results are shown as a way of comparing the two controllers.

The controller shown in Figure~\ref{fig:model1} is implemented as routine running single-threaded on one dedicated core. Changes to the control algorithm itself appear as matrix operations whose coefficients are calculated prior to the experiment (system identification model and control design) and fixed during the experiment.

\subsection{ $F$ control results}
\label{sec4a}
The $F$ feedback controller is implemented in the MCS and tested on MST. The control time occurs between $t_1 = 0.02$~s and  $t_2 = 0.05$~s where we choose to track square oscillations. Before time $t_1$ we are in an open loop mode. At time $t_1$ we activate the control mode through the MCS and at time $t_2$ we release the system to open loop again. 
\begin{figure} 
\centering
\includegraphics [width=1\linewidth]{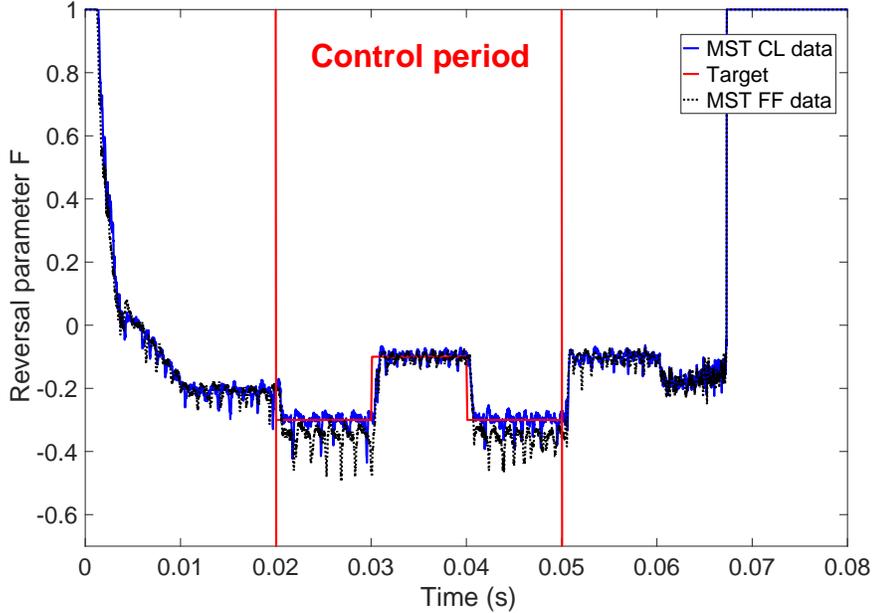} 
\caption{Comparison of outputs of $F$ control with and without feedback control during a tracking task.}
\label{Fcontrol}
\end{figure}
Figure~\ref{Fcontrol} represents the output response of MST through $F$ measurements with and without feedback control.
As expected, the time-dependent results of the closed loop experiments successfully track the target during the control period. This tracking is better if compared to the open loop case (dashed line) where steady state errors appear. It is important to notice that despite the control effect, in both cases, sawtooth crashes are still occurring throughout the run.

Note that for many operational purposes in the RFP, open-loop, feed-forward control is adequate, since waveforms can be optimized empirically, shot-to-shot.  Often, however, particularly in situations where the plasma response is both important and difficult or inconvenient to predict in advance, closed-loop feedback control is needed.  In the context of $F$ control, although it is possible to tune a pre-programmed $B_T$ waveform shot-to-shot to approximately achieve the desired waveform, as in the black signal in Figure~\ref{Fcontrol}, shot-to-shot changes in the time evolution of $B_T$ still affect the ratio $F$, and better control can be achieved with feedback on the real-time signals, as discussed above.  The superiority of closed-loop over open-loop control is expected to be especially important in the cases of OFCD~\cite{McCollam10}, PPCD~\cite{Chapman01}, and SSRD~\cite{Nebel02}.

\subsection{ $I_\text{p}$ control results}
\begin{figure} 
\centering
\includegraphics [width=1\linewidth]{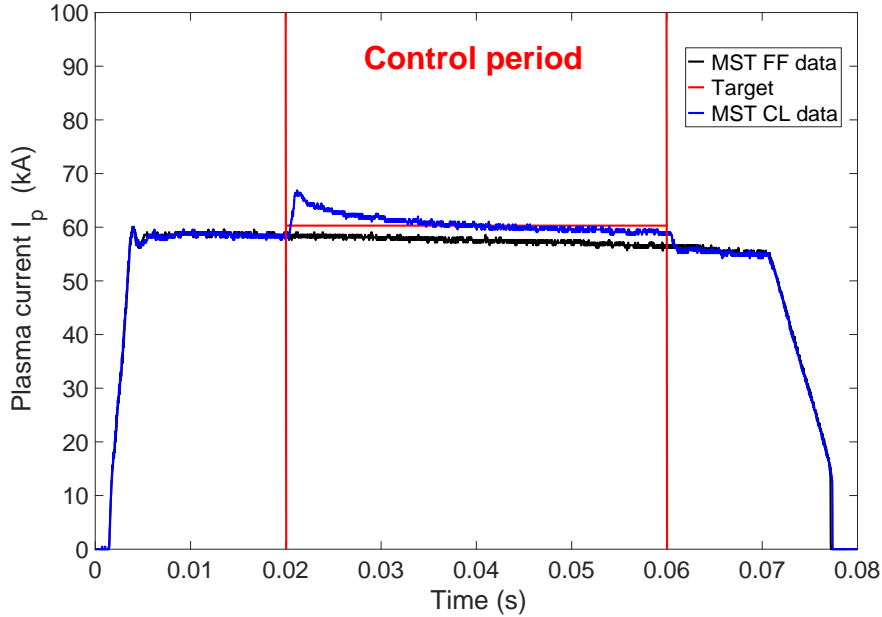} 
\caption{Comparison of outputs of $I_\text{p}$ measurements with and without feedback control during a flat tracking.}
\label{IPcontrol1}
\end{figure}

The $I_\text{p}$ controller is implemented in the MCS and tested on MST. The control time occurs between $t_1 = 0.02$~s and  $t_2 = 0.06$~s. We chose to take a longer control time frame so we can observe the plasma resistivity effects that act as a drag in the plasma current value towards the end of the shots.
Figure~\ref{IPcontrol1} represents the output response of MST through $I_\text{p}$ measurements with and without control.
The time-dependent results of the closed loop experiment successfully track the flat target during the control period. This tracking is better if compared to the open loop (FF) case (black line) where an important steady state error appears. The plasma resistance add more steady state error to both cases towards the end, but the controller does a better tracking despite the drag. 

\begin{figure} 
\centering
\includegraphics [width=1\linewidth]{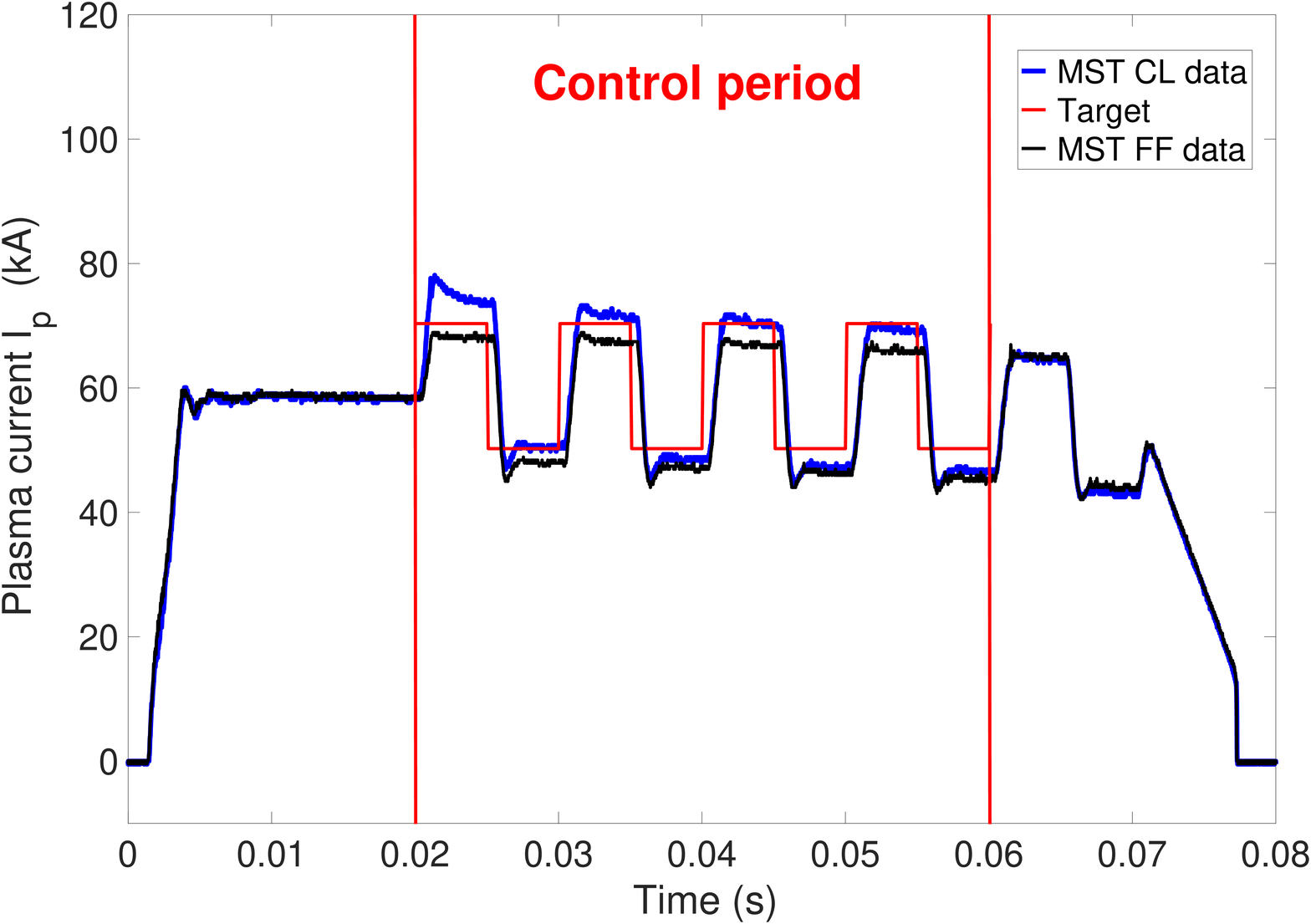} 
\caption{Comparison of outputs of $I_\text{p}$ measurements with and without feedback control (a) during a square wave tracking and its corresponding input ($I_{pg}$) (b).}
\label{IPcontrol2}
\end{figure}

\begin{figure} 
\centering
\includegraphics [width=1\linewidth]{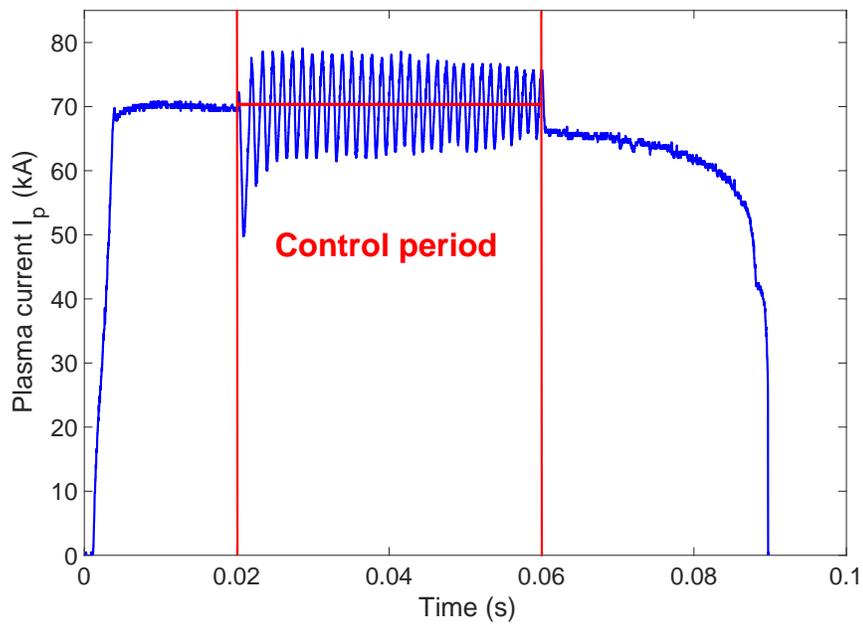} 
\caption{Resulting $I_\text{p}$ measurements with feedback control during a flat tracking}
\label{IPcontrol3}
\end{figure}

Figure~\ref{IPcontrol2} is similar to Figure~\ref{IPcontrol1} but for a different target: a square wave. We notice the same observations as before where there is a little overshoot at the beginning of the control period that dissipates slowly through the shot until the plasma resistance becomes important enough to start dragging the current down. One can argue that the controller is not aggressive enough to overcome this drag or not fast enough to get to the target. The tuning of the LQE control gains is indeed critical in this case but we have found that increasing the integrator or the feedback gain too much results in disruption of the plasma or introduces oscillations that we do want to avoid. We found some trade off values that would allow us to get to the target in less than $20$~ms and minimize the steady state error. Figure~\ref{IPcontrol3} is an example where we emphasized the importance of zero steady state error at the expense of the system stability. In this example we pushed the controller (integrator) to its limits so we can beat the plasma resistivity. The resulted plasma current was extremely oscillatory which cannot be considered a possible solution of the controller.

\subsection{ Coupling control results}
For the two inputs two outputs control results, we studied two independently designed controllers for $F$ and $I_\text{p}$. The first case is called parallel control in the MCS where we connected the two independent controllers designed from the two independent models of $F$ and $I_\text{p}$ in parallel without sharing any knowledge between each other. The second case is the coupled control case where the controller is designed directly from the coupled system of $F$ and $I_\text{p}$.

\begin{figure} 
\centering
\includegraphics [width=1\linewidth]{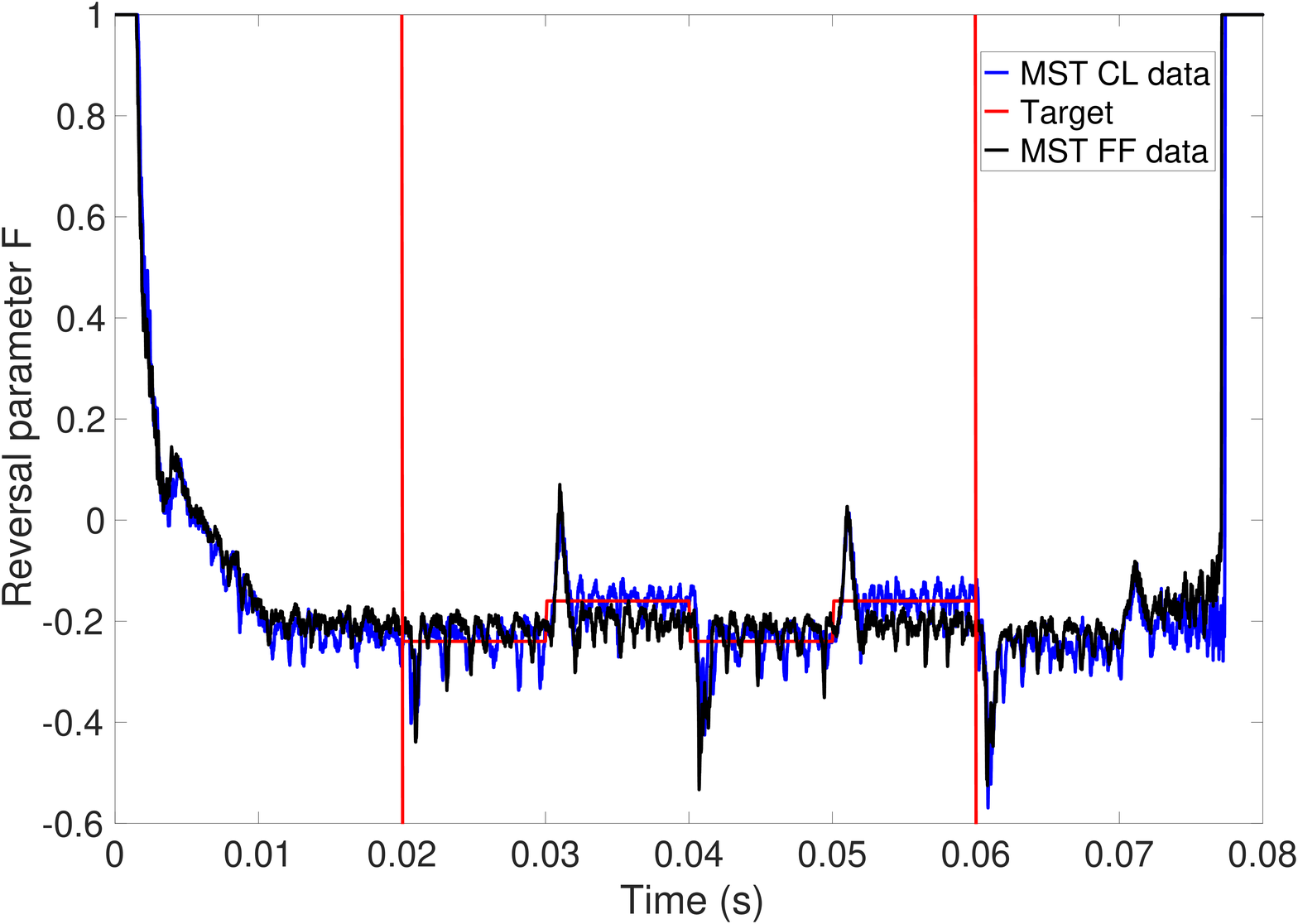}  \hspace{-5.5em}\raisebox{10em}{(a)} \\
\includegraphics [width=1\linewidth]{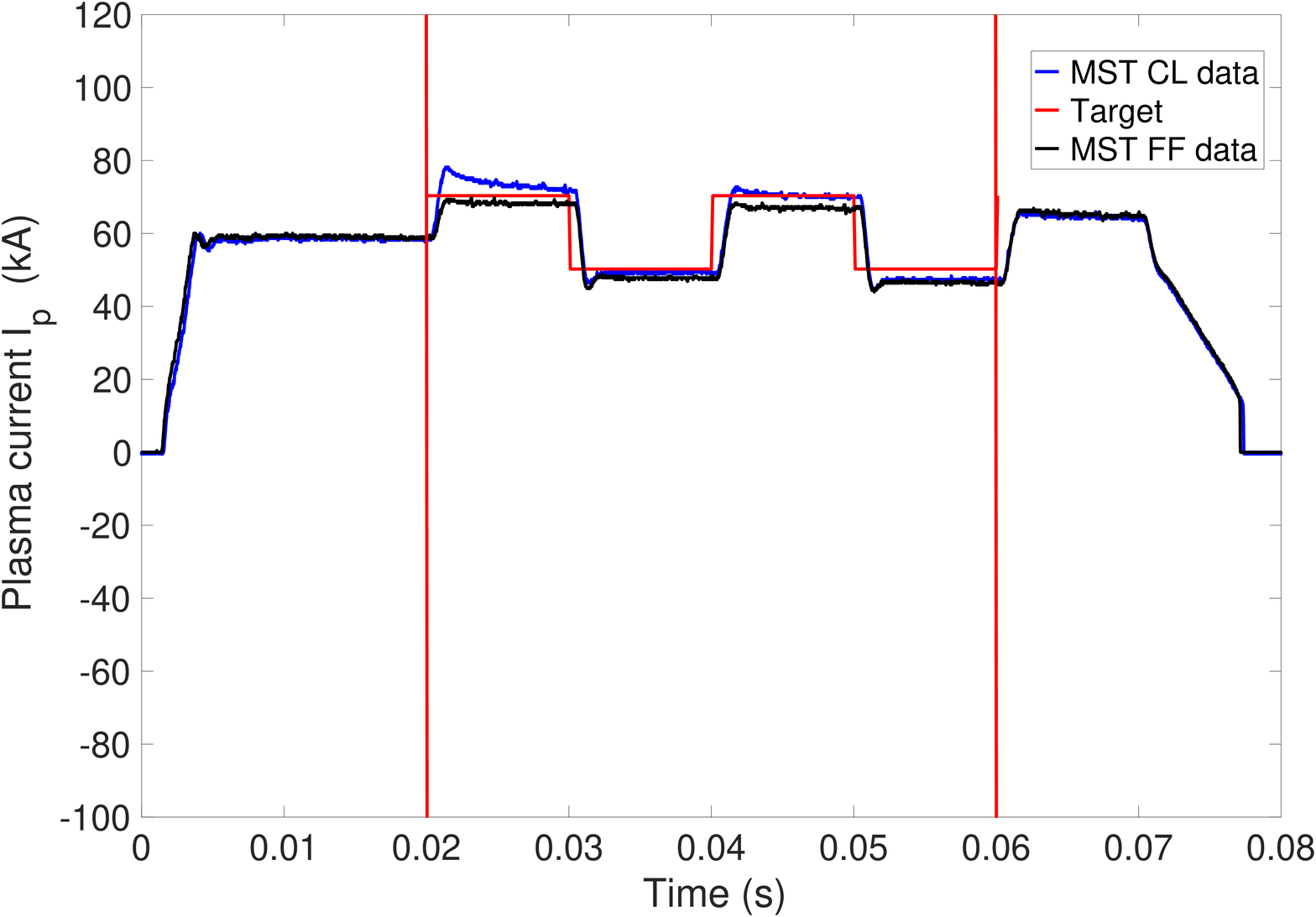}  \hspace{-5.5em}\raisebox{8.7em}{(b)} 
\caption{Comparison of outputs of $F$ and $I_\text{p}$ measurements with and without parallel feedback control during a square wave tracking.}
\label{Parallel}
\end{figure}

Figure~\ref{Parallel} shows an example of parallel control results where we compare measurements of reversal parameter $F$ and plasma current $I_\text{p}$ when we track a square wave using the double closed loops controller and the feedforward controller between the times $t_1 = 0.02$~s and $t_2 = 0.06$~s. We notice that despite the overshoot, we are able to successfully track the wave in both $F$ and $I_\text{p}$ outputs.

\begin{figure} 
\centering
\includegraphics [width=1\linewidth]{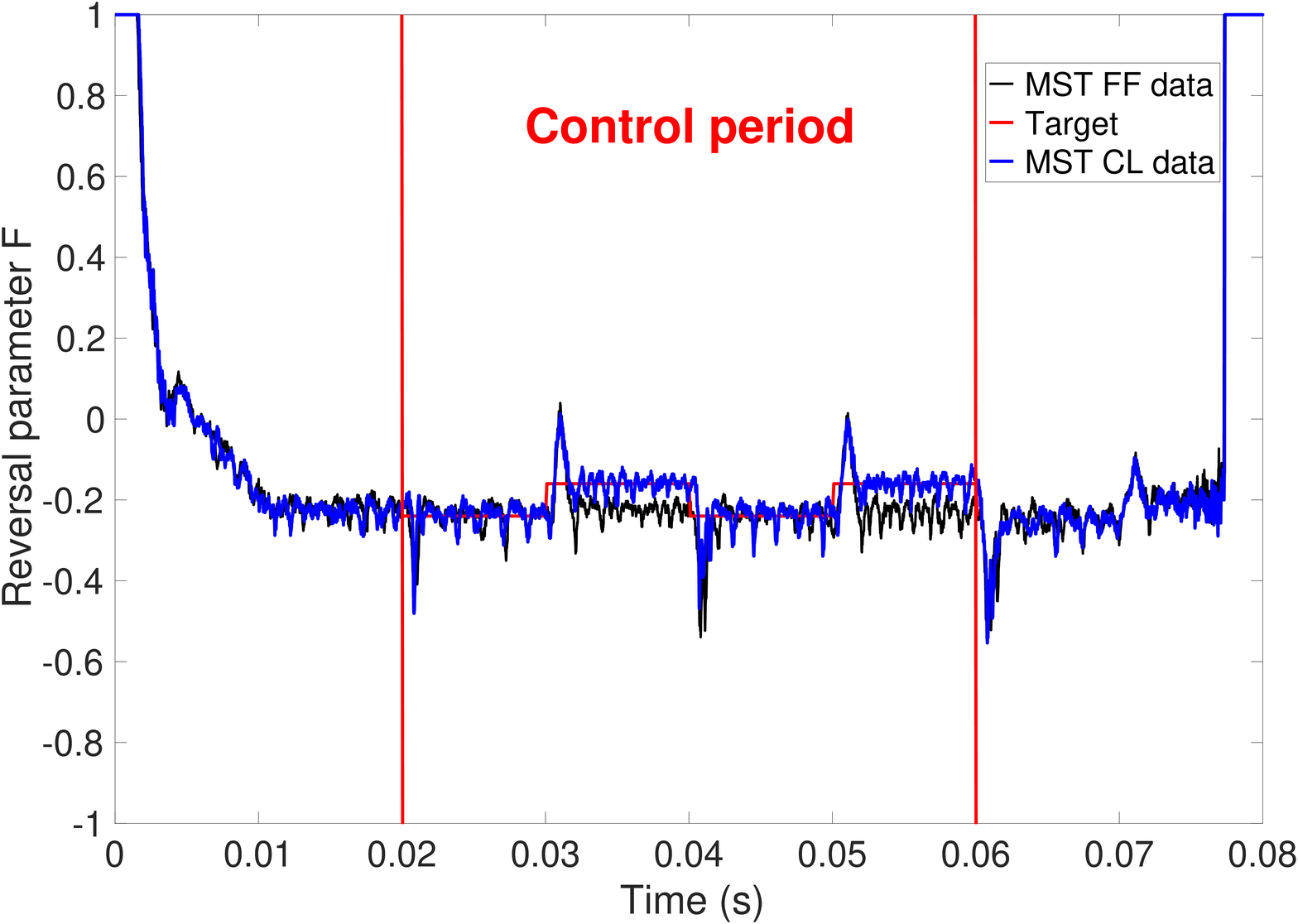}  \hspace{-5.5em}\raisebox{10em}{(a)} \\
\includegraphics [width=1\linewidth]{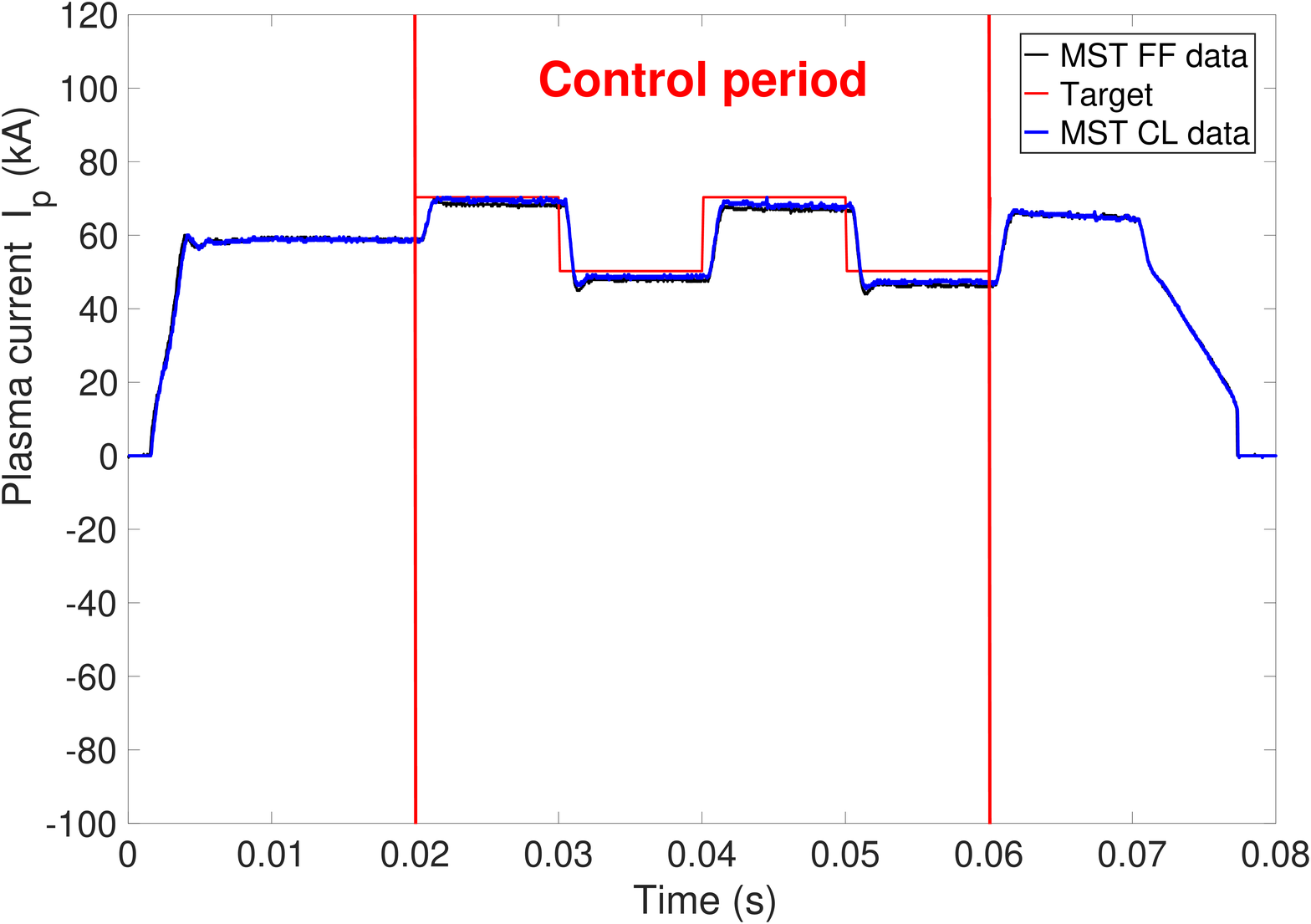}  \hspace{-5.5em}\raisebox{8.7em}{(b)} 
\caption{Comparison of outputs of $F$ and $I_\text{p}$ measurements with and without Coupled feedback control during a square wave tracking.}
\label{Coupled1}
\end{figure}

Designing the coupled model had a goal of improving the controller by giving it access to the coupling dynamics. Figure~\ref{Coupled1} shows a different example of coupled control results where we compare measurements of reversal parameter $F$ and plasma current $I_\text{p}$ when we track the same square wave using the closed loop controller and the feedforward controller between the same times $t_1 = 0.02$~s and $t_2 = 0.06$~s. We can notice a slight improvement in the $F$ control but
also
an overshoot elimination in the plasma current $I_\text{p}$ control. This feature is
important in our experiments, in that we are presently restricted by an IGBT current limit to 80~kA of maximum plasma current.

\subsection{Discussion}

From the experimental results shown in this paper, good performance was obtained using either the MIMO optimal controller or two SISO loop control design. From the perspective of an operator, this is desirable, as we showed that the two loop structure, with a small number of free parameters that can be adjusted intuitively between shots, works well in experiments.
However, the optimal design provides a more systematic algorithm for designing a stabilizing controller. It is well suited for handling systems with strong cross-coupling, and can be easily extended to include additional controlled variables and actuators. As scenarios that exhibit stronger coupling are explored, or as additional actuators and controlled outputs are considered, the tuning of separate PID loops will become more difficult, while the MIMO control design approach will still be appropriate.

As mentioned in Section~\ref{intro},
classical, physics-based approaches have been taken for controlling $I_p$~\cite{Bettini11} and $F$~\cite{Barp11} in the RFX-mod RFP, using linear controllers with advanced feedforward or PIDs adjusted via pole placement based on transfer functions derived from simplified physical models of the plasma dynamics.
While it is too early for precise comparisons between that work and our system-identification approach, we can nonetheless compare the merits of the approaches in a more general context.
Physics-based models have the advantage of being immediately interpretable as the states of the system have meaningful physical representations.
This desirable property makes the manual tuning of parameters easier and can provide insights into the science.
On the other hand, system identification, while more opaque, has the benefit of being based on machine response rather than presumed physics models, and providing some freedom on how accurately to capture the dynamics by choosing the size of the state vector.
And modern MIMO controllers offer systematic ways of designing robust stabilizing controllers with any number of sensors and actuators without necessitating as much manual tuning.

\section{Conclusion}
\label{sec5}
A novel system has been implemented at MST to systematically control $F$ and $I_\text{p}$ individually or simultaneously in RFP plasmas via the two primary current actuators $I_{pg}$ and $I_{tg}$. Initial experiments with the closed feedback control loop show promising results and motivate future work of continued testing and design improvements. In addition, manipulation of $I_{pg}$ and $I_{tg}$ can be integrated into a more complex control scheme, for example by including cycle-averaged plasma current or loop voltage as controlled quantities for cases with OFCD.  

As part of this work, a flexible framework for performing feedback control design and experimentation on the MST has been developed. This framework will aid in the creation of advanced control algorithms by providing means for conducting system identification simulations and high-fidelity tests of proposed algorithms prior to and during experimental implementation and testing. In the longer term, the same framework could be extended to include additional actuators and measurements on MST, including density control through gas puffing or loop voltage control based on magnetic fluctuation amplitudes.

These methodologies are a key element of research toward advanced inductive control of an ohmically heated RFP fusion plasma.

\section*{Acknowledgement}
The authors acknowledge helpful discussions with Dr. Brett Chapman.
This material is based upon work supported by the U.S. Department of Energy Office of Science, Office of Fusion Energy Sciences program under Award Numbers DE-FC02-05ER54814 and DE-SC0018266.\\
Part of this work was performed under the auspices of the U.S. Department of Energy by Lawrence Livermore National Laboratory under Contract DE-AC52-07NA27344.\\

\bibliographystyle{unsrt}
\bibliography{Control_paper_2020}

\end{document}